\documentclass[twocolumn,amsmath,amssymb,prl,aps,showkeys]{revtex4}
\usepackage{graphicx}
\usepackage{amsmath,amssymb}
\usepackage{mathrsfs}
\usepackage{color}
\usepackage{afterpage}
\usepackage[version=3]{mhchem}
\usepackage[normal]{subfigure}
\usepackage{natbib}

\begin{document}
\title{Exploring N-rich phases in Li$_x$N$_y$ clusters for hydrogen storage at nano-scale}
\author{Amrita Bhattacharya$^1$, Saswata Bhattacharya$^{2*}$} 
\affiliation{$^1$Fritz-Haber-Institut der Max-Planck-Gesellschaft, Faradayweg 4-6, D-14195 Berlin, Germany \\ $^2$Department of Physics, Indian Institute of Technology Delhi, Hauz Khas 110016, New Delhi, India}
\date{\today}
\begin{abstract}
We have performed cascade genetic algorithm and {\em ab initio} atomistic thermodynamics under the framework of first-principles density functional theory to study the (meta-)stability of a wide range of Li$_x$N$_y$ clusters. We found that hybrid xc-functional is essential to address this problem as a local/semi-local functional simply fails even to predict a qualitative prediction. Most importantly, we find that though in bulk Lithium Nitride, Li rich phase, i.e. Li$_3$N, is the stable stoichiometry, in small Li$_x$N$_y$ clusters N-rich phases are more stable at thermodynamic equilibrium. We further show a that these N-rich clusters are promising hydrogen storage material because of their easy adsorption and desorption ability at respectively low ($\leq$ 300K) and moderately high temperature ($\geq$ 600K). 
\end{abstract}
\maketitle
\noindent Hydrogen is the most abundant element in the universe that contains the highest energy density per unit mass, and when it burns it produces only water and energy. In view of this, it has been considered to be one of the promising solutions for the clean alternative energies~\cite{Zuttel2003, Schalapbach2001,Jena2011}. Significant amount of research initiatives have been taken place along this direction~\cite{Orimo2007, Suh2011, Murray2009, sas-licanh}, but unfortunately, hydrogen is needed to be produced from water or other solid-state materials as it is not freely available in nature. Therefore, although hydrogen can be considered as renewable, if it is produced from water, it costs more energy to get produced, than the energy one recovers on burning it. This necessitates, designing of promising solid-state materials or nano-structures, where hydrogen molecules can be stored. This has been, therefore, an active field of research for the last few decades and various kind of materials such as metal hydrides~\cite{Orimo2007, Sakintuna2007, Jain2010}, clusters~\cite{sas-liab, Jose2011, Watari2000, Wagemans2005, Sun2005, Sun2006, Li2009, Koukaras2012}, nanostructures~\cite{Norberg2011, Wang2009, Arico2005, sas-kubas,sas-bcn1, sas-bcn2, amritatife}, highly porous metal-organic frameworks (MOFs)~\cite{Rowsell2005, Suh2011, Murray2009, Kesanli2005, Wong-Foy2006, Li1999}, clathrate hydrates~\cite{Florusse2004, Lee2005, Chapoy2007}, covalent organic frameworks (COFs)~\cite{Han2008, Klontzas2008}  etc. have been studied extensively. In particular, complex binary hydrides involving light metals such as Li, Mg, Ca etc. have been extensively investigated because of their high gravimetric storage capacity. Amongst them, lithium nitride [Li$_3$N], lithium imide [Li$_2$NH] or amide[LiNH$_2$] have been found to exhibit strong affinity for H$_2$ because of the reversible reactions Li$_3$N+2H$_2$ $\leftrightarrow$ Li$_2$NH+LiH+H$_2$ $\leftrightarrow$ LiNH$_2$ +2LiH~\cite{chen2002}. Unfortunately, though this material can store sufficiently high amount of hydrogen, the desorption kinetics is poor for its on-board practical application. Li atoms being ionized as Li$^+$ cations, strongly attracts [NH$_2$]$^-$ complex anion and thereby form a strong bond, making the desorption of hydrogen to be high~\cite{sas-licanh}. In order to lower the desorption temperature to a permissible range, doping suitable metal ad-atoms have already been tried~\cite{ahuja2007,sas-licanh}. 

\noindent We have explored the possibility of realising clusters of these materials. It has been found that reducing the number of particles in a cluster reveals the possibility of several interesting properties. In a range where matter is reduced to sizes of only a few atoms, the intrinsic properties of the so-called clusters are non-scalable from their bulk analogues. We expect that significant amount of hydrogen adsorption would be possible in a non-dissociative manner due to presence of high surface-to-bulk ratio and due to the weak dispersive interactions the hydrogen desorption could be favourable. In past, it's been studied to understand the hydrogen storage efficiency of small lithium amide clusters [LiNH$_2$]$_n$ using {\em ab initio} molecular orbital theory~\cite{Lanlan2010}. But materials property changes under operational environment (e.g. temperature ($T$) and pressure ($p$) in an atmosphere of reactive molecules). The thermodynamics and kinetics at the relevant temperature ($T$) and the nature of the environment determine the composition and structure of clusters. In thermodynamic equilibrium, only structures and compositions that minimize the Gibbs' free energy of formation of the composite cluster+ligands(gas) system will be the most stable. Therefore, one has to first ensure the most stable phases and compositions of such clusters at thermodynamic equilibrium. And after that the efficiency of hydrogen storage should be aimed at. Till date, no such consistent theoretical study has been performed, in the field of hydrogen storage in small clusters, that has considered the (meta-)stability of the clusters by including the effect of temperature and pressure of the reactive environment. In this letter, we address the issue of stability and metastability using our model system that has significance in many practical applications. Our model includes free metal (Li) clusters in a (a) nitrogen only atmosphere and then in (b) an atmosphere of both nitrogen and hydrogen. It should be noted that we have addressed the situation under thermodynamic equilibrium, while the real system might not be always in thermodynamic equilibrium. But we emphasize that, a thermodynamic phase diagram acts as guideline and depicts its limit for predicting properties and functions of real materials. 

\noindent We have considered first a wide range of Li$_x$N$_y$ clusters, where $x$=1, 2,..., 6, 9, 12, 15, 18 and $y$ is determined by thermal equilibrium with the environment at given temperature $T$ and partial nitrogen pressure $p_{\textrm N_2}$\footnote{We varied $y$ value starting from 1, 2, 3,... and kept on increasing it until when the specific Li$_x$N$_y$ stoichiometry goes totally outside our phase diagrams, i.e. under no circumstances that stoichiometry can be important in the environmental conditions which we have considered here.}. To get the minimum energy configurations, for each stoichiometry, the total energy is minimized with respect to both geometry and spin state. Very unexpectedly, our results reveal that though in bulk Li$_3$N is the most stable phase, in small clusters, the stoichiometry of the most stable phase is quite different.

\noindent The low energy structures (including the global minimum~(GM)) are generated from an exhaustive scanning of the potential energy surface~(PES) from our recent implementation of cascade genetic algorithm. The term cascade means a multi-stepped algorithm where successive steps employ higher level of theory and each of the next level takes information obtained at its immediate lower level. Typically the cascade GA starts from a crude but sophisticated classical force field and goes up to density functional theory using hybrid functionals. This GA algorithm's implementation is thoroughly benchmarked and it's efficiency is validated (w.r.t more advanced theory) in details in Ref~\cite{Bhattacharya2013, Bhattacharya2014}.
\begin{figure}[t!]
\includegraphics[width=1.0\columnwidth,clip]{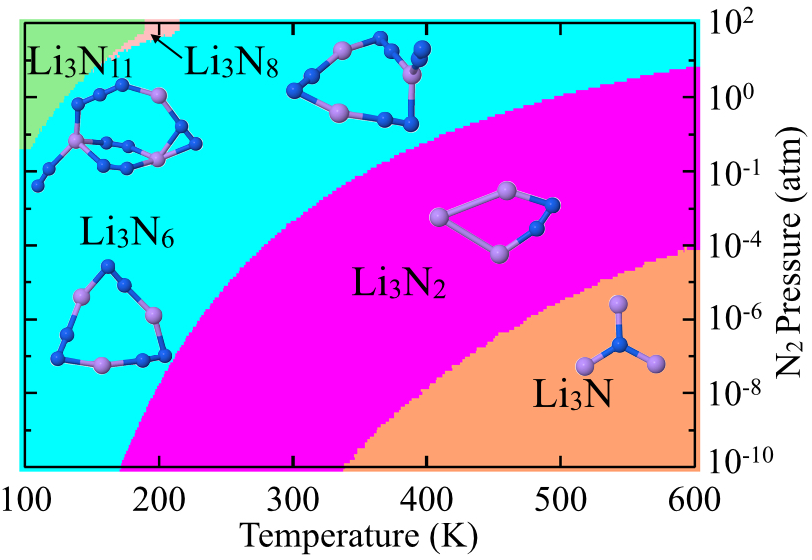}
\caption{Phase diagram for Li$_3$N$_x$ clusters in an nitrogen atmosphere. The geometries are optimized with PBE+vdW and the total energies are calculated with PBE0+vdW.}
\label{fig2}
\end{figure} 

\noindent We have performed the density functional theory (DFT) calculations using FHI-aims, which is an all electron code with numerical atom centred basis sets~\cite{Blum2008}. The low energy GA structures are further optimized at a higher level settings, where energy minimization is performed with vdW-corrected-PBE+vdW~\cite{Perdew1997} functional, ``tight - tier 2" settings, and force tolerance was set to better than 10$^{-5}$ eV/${\textrm \AA}$.  The van der Walls correction is calculated as implemented in Tkatchenko-Scheffler scheme~\cite{Tkatchenko2009}. Finally the total single point energy is calculated afterwards on top of this optimized structure via vdW-corrected-PBE0~\cite{Perdew1996} hybrid xc functional (PBE0+vdW), with ``tight - tier 2" settings. \footnote {As described in details in Ref~\cite{Bhattacharya2014}, in our cascade GA, the latter energy is used to evaluate the fitness function, i.e., a mapping of the energy interval between highest-and lowest-energy cluster in the running pool into the interval [0, 1]. Obviously, the higher the value of the fitness function for a cluster, the higher is the probability of selecting it for generating a new structure.} We find that PBE+vdW strongly overestimates stability of clusters with larger $y$ values in Li$_x$N$_y$ clusters. This results a qualitatively wrong prediction that adsorption of N$_2$  could be favored over desorption up to a large excess of nitrogen. Such behavior is not confirmed by hybrid functional [e.g HSE06~\cite{Heyd2006}, PBE0] as employed in our calculations. The difference in energetics of PBE0 and HSE06 is always within 0.05 eV. The spin states of the clusters are also different as found by PBE and PBE0/HSE06. In view of this, all our results are thoroughly tested and benchmarked w.r.t hybrid functionals (PBE0) using ``tight" numerical settings and tier 2 basis set~\cite{Blum2008}. 

\noindent The free energy of the low energy isomers\footnote{we have considered all the isomers within an energy window of 0.5 eV from the global minimum (GM) as we have seen that it's very unlikely that isomers above 0.5 eV from the GM would become more stable after including their translational, rotational, vibrational, spin and symmetry free energy contributions to the total energy.} (in the PES) is then calculated as a function of $T$ and $p_{\textrm N_2}$ for each stoichiometry using the {\em ab initio} atomistic thermodynamics (aiAT) approach. The concept of aiAT is earlier developed and successfully applied  initially for bulk semiconductors~\cite{scheffler1986}, and later applied to the study of oxide formation at the surface of some transition metals and other materials~\cite{reuter-aiat}. We have recently extended this approach to cluster systems~\cite{Bhattacharya2013} following our detailed description as in Ref \cite{Bhattacharya2014}. Therefore, from different cluster compositions and structures with the lowest free energy the thermodynamic phase diagram can be constructed as a function of $T$ and $p_{\textrm N_2}$. One such phase diagram is shown in Fig.~\ref{fig2}. At a given $T$, $p_{{\textrm N}_2}$, and $y$, the stable stoichiometry of a Li$_x$N$_y$ cluster is determined via aiAT, i.e., by minimizing the Gibbs' free energy of formation $\Delta G_f (T,p_{{\textrm N}_2})$.
\begin{equation}
\Delta G_f (T,p_{{\textrm N}_2}) = F_{{\textrm Li}_x{\textrm N}_y}(T) -  F_{{\textrm Li}_x}(T) - y\times\mu_\textrm{N}(T,p_{{\textrm N}_2})
\label{eq1}
\end{equation}

\begin{figure}[t!]
\includegraphics[width=1\columnwidth,clip]{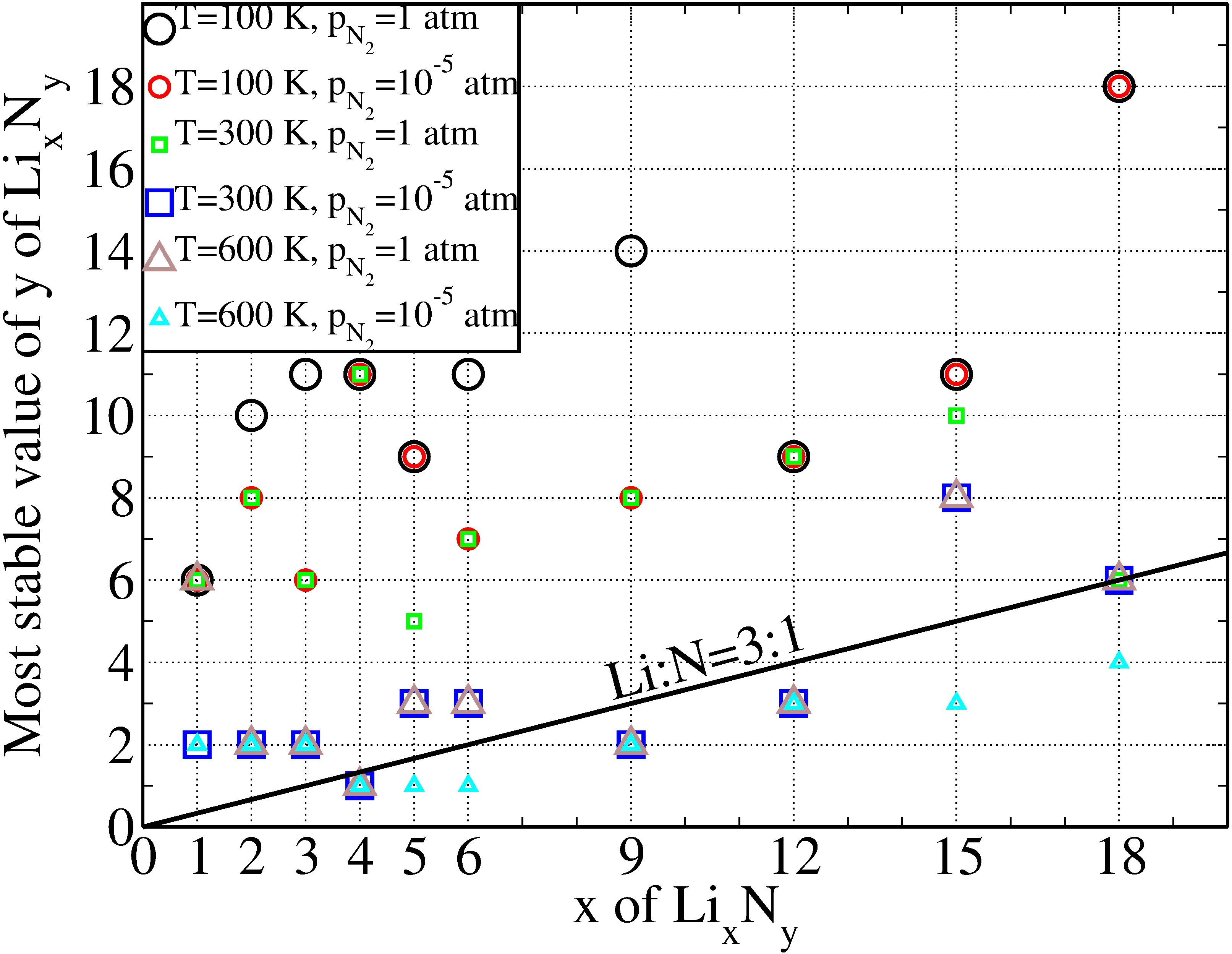}
\caption{(color online) The most stable Li$_x$N$_y$ clusters at various temperatures and pressures under thermodynamic equilibrium. The geometries are optimized with PBE+vdW, and the electronic energy is calculated using PBE0+vdW.}
\label{fig1}
\end{figure} 
\noindent Here, $F_{{\textrm Li}_x{\textrm N}_y}(T)$ and $F_{{\textrm Li}_x}(T)$ are the Helmholtz free energies of the Li$_x$N$_y$ and the pristine Li$_x$ cluster\footnote {The clusters are at their ground state configuration w. r. t. geometry and spin, respectively.} and $\mu_\textrm{N}(T,p_{{\textrm N}_2})$ is the chemical potential of nitrogen. As explained in Ref~\cite{Bhattacharya2014}, $F_{{\textrm Li}_x{\textrm N}_y}(T)$ and $F_{{\textrm Li}_x}(T)$ are calculated using DFT information and are calculated from the sum of DFT total energy, DFT vibrational free energy in the harmonic approximation, as well as translational, rotational, symmetry and spin-degeneracy free-energy contributions. The dependence of $\mu_\textrm{N}(T,p_{{\textrm N}_2})$ on $T$ and $p_{{\textrm N}_2}$ is calculated using the ideal (diatomic) gas approximation with the same DFT functional as for the clusters. The phase diagram for a particular Li$_x$N$_y$ is constructed by identifying the lowest free-energy structures at each $T$, $p_{\textrm N_2}$. As a representative example, we show in Fig.~\ref{fig2} the phase diagram for $x$ = 3. Note that surprisingly in small cluster, we find that at a realistic $T$ and $p_{{\textrm N}_2}$, N-rich phases are becoming more stable over the conventionally known most stable phase Li$_3$N as in bulk. We have then extended our observation for a wide range of clusters of Li$_x$N$_y$ configuration varying $x$ as 1, 2,.., 6, 9, 12, 15, 18 and $y$ is increased as 1, 2, 3,.. as per the highest possible value at each $x$ in the thermodynamic condition as employed in the phase diagrams. For example, for $x=3$, we see from Fig.~\ref{fig2} the highest possible $y$ is 11. We find that the N-rich phase in these small clusters are consistently becoming quite stable over a wide range of ($T, p_{{\textrm N}_2}$) environmental conditions. This is shown in Fig.~\ref{fig1} where we have plotted the most stable stoichiometry of $y$ at different ($T, p_{{\textrm N}_2}$) with varying $x$ values. The black line represents the usual Li:N=3:1 stoichiometry as in bulk. Note that most of the stable phases are above this black line and are significantly high N-rich especially for small clusters. As expected, due to quantum size effect, the reactivity of the larger clusters get decreased with growing size. But still even at larger sizes, the clusters are more N-rich as compared to their bulk limit. Only at high temperature ($>$ 600 K) and low pressure ($<$ 10$^{-5}$ atm) the clusters are N-deficient.
\begin{figure}[h!]
\includegraphics[width=1.0\columnwidth,clip]{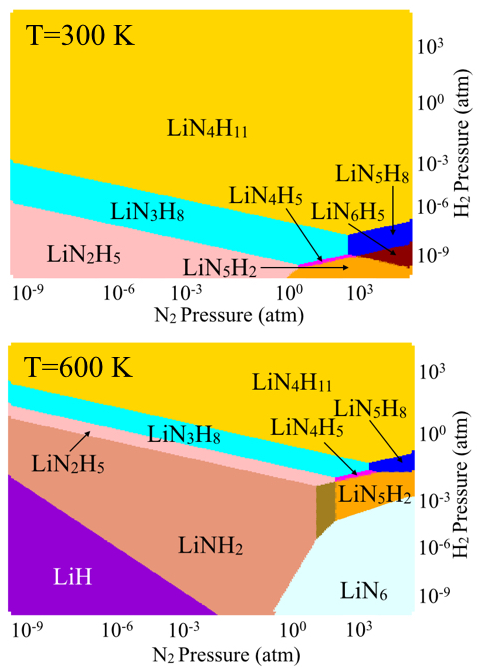}
\caption{(color online) Phase diagram of LiN$_y$H$_z$ at varying pressure of $p_{{\textrm H}_2}$ and $p_{{\textrm N}_2}$ of two different temperatures T = 300 K (top) and T = 600 K (bottom)}
\label{fig3}
\end{figure}  
 
\noindent Subsequently, we explored the efficiency of the N-rich phases of these cluster as hydrogen storage material. 
However, the procedure for this is not easy as one has to understand under which temperature conditions and in which N$_2$ and H$_2$ (partial) pressure range, these nanoclusters become more stable, i.e. under which optimum conditions the N$_2$ and H$_2$ adsorption can take place on the surface of a given cluster size.  Note that under reaction conditions, the nano-cluster comprises of a wide range of structures including different number of atoms with various oxidation states, all of which could be active to some extent in the reaction. Some obvious questions arise naturally, for example, ``which are the species present in the real reaction and what are their structures?". It's also interesting to understand ``how the nano-cluster changes their structure and properties upon adsorption of a different ligand molecules? We have addressed these question on considering another ligand (e.g H$_2$) adsorbed onto LiN$_y$ clusters forming LiN$_y$H$_z$ clusters and tried to understand the stability of the entire system in presence of both N$_2$+H$_2$ pressures. 
In order to fulfil these aims, small clusters LiN$_y$H$_z$ are first thoroughly scanned from our cascade genetic algorithm implementation and their stability is studied using atomistic thermodynamics to mimic an atmosphere composed of N$_2$ and H$_2$ gases at realistic temperature and pressure conditions. As depicted in Eq.~\ref{eq1}, here to address the (meta-)stability of LiN$_y$H$_z$ clusters, in presence of one more additional ligands i.e. H$_2$, the equation gets modified as below: 

\begin{equation}
\begin{split}
\Delta G_f (T,p_{{\textrm N}_2},p_{{\textrm H}_2}) = F_{{\textrm Li}{\textrm N}_y{\textrm H}_z}(T) -  F_{{\textrm Li}}(T) \\ - y\times\mu_\textrm{N}(T,p_{{\textrm N}_2}) - z\times\mu_\textrm{H}(T,p_{{\textrm H}_2})
\end{split}
\label{eq2}
\end{equation}

\noindent Here, $F_{{\textrm Li}{\textrm N}_y{\textrm H}_z}(T)$ and $F_{{\textrm Li}}(T)$ are the Helmholtz free energies of the LiN$_y$H$_z$ and the pristine Li cluster (in this case single Li atom) respectively. Like before as in Eq.~\ref{eq1}, here also we have calculated this from the sum of PBE0 DFT total energy, DFT vibrational free energy (in the harmonic approximation), and translational, rotational, symmetry, spin-degeneracy free-energy contributions. $\mu_N(T,p_{{\textrm N}_2})$ and $\mu_\textrm{H}(T,p_{{\textrm H}_2})$ are the chemical potential of nitrogen and hydrogen respectively. 

\noindent In Fig.~\ref{fig3}, we have compared the phase diagrams of the stability of LiN$_y$H$_z$ clusters, varying H$_2$ and N$_2$ pressures, at two different temperatures (viz. 300 K (top) and 600 K (bottom)). It's clear from Fig.~\ref{fig3} that at T=300 K, the small clusters can adsorb quite a substantial amount of hydrogen at a realistic temperature and pressure. Note that in bulk Li$_3$N, the highest amount of hydrogen uptake is possible when it forms lithium amides/imides (LiNH$_2$/Li$_2$NH). But in small clusters, the situation is very different as LiN$_4$H$_{11}$ is thermodynamically the most stable cluster and the material is stable over a wide experimentally achievable region in the phase diagram at $T$=300 K.
\begin{figure}[t!]
\includegraphics[width=0.90\columnwidth,clip]{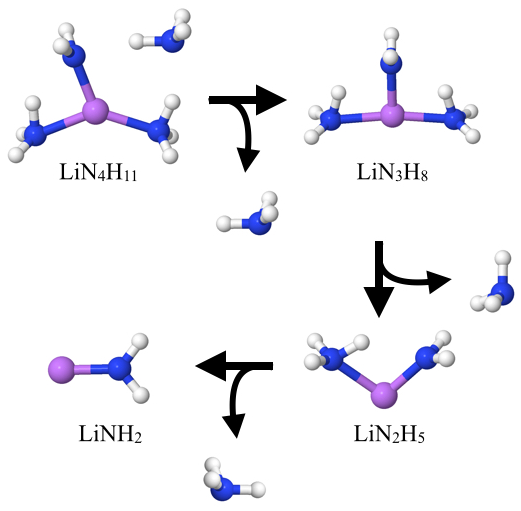}
\caption{(color online) Release of three units of ammonia from LiN$_4$H$_{11}$ before the final product LiNH$_2$ is formed at a varying $T$, $p_{{\textrm N}_2}$, $p_{{\textrm H}_2}$.}
\label{fig4}
\end{figure}  
\noindent It should be noted that for an efficient hydrogen storage materials, the desorption kinetics should also needs to be optimal. We have provided a similar phases diagram in the bottom panel of Fig.~\ref{fig3} at T = 600 K. We find that at this temperature first the stability region of LiN$_4$H$_{11}$ is reduced, and less H-rich phases (viz. LiN$_3$H$_8$, LiN$_2$H$_5$, LiNH$_2$) become more stable. This means the cluster tends to leave hydrogen at higher temperature, while the H-rich phase being absolutely stable at room temperature. On a critical look to the phase diagram, it appears to be that under a fixed pressure region of both $p_{{\textrm H}_2}$ and $p_{{\textrm N}_2}$, if we simply increase temperature these type of clusters have a tendency to release ammonia (NH$_3$) molecule. This is shown more clearly in Fig.~\ref{fig4}. It's evident from Fig.~\ref{fig3} that on increasing temperature these nano-clusters will form a reversible cycle under a constant $p_{{\textrm H}_2}$ and $p_{{\textrm N}_2}$, where the following reaction is going on as shown in Fig.~\ref{fig4}:
\begin{equation}
\textrm {LiN}_4{\textrm H}_{11} \rightarrow \textrm {LiN}_3{\textrm H}_{8} + \textrm{NH}_3 \rightarrow  \textrm {LiN}_2{\textrm H}_{5} + \textrm{NH}_3 \rightarrow \textrm {LiN}{\textrm H}_{2} + \textrm{NH}_3
\end{equation}
Clearly, just by changing temperature we can get three units of NH$_3$ out of this LiN$_4$H$_{11}$. Therefore, getting such an automatic release of ammonia solves the problem of hydrogen storage, as hydrogen can be released economically from ammonia on-demand, without the need for high-pressure or cryogenic storage~\cite{Lan2012}.

\noindent In summary, we have employed our massively parallel cascade genetic algorithm to scan the potential energy surface of a large number of Li$_x$N$_y$ clusters. We find that a local/semi-local functional is not correct even to find a reasonably qualitative prediction. In view of this, we have employed more advanced hybrid DFT functional PBE0 to calculate accurately the energetics throughout our calculations. We find by applying aiAT, that the behaviour of small Li$_x$N$_y$ clusters are very different than their bulk, which may be due to quantum confinement effect. In small clusters, over a wide range of sizes, N-rich phases are consistently stable at thermodynamic equilibrium under experimentally achievable temperature and pressure $p_{{\textrm N}_2}$. We, therefore, presumably for the first time introduce a new class of clusters, which has been overlooked in the past. We went one step further to understand the effectiveness of these clusters in the field of hydrogen storage. We generated a large number of Li$_x$N$_y$H$_z$ clusters out of which one of the test cases is shown here for LiN$_y$H$_z$. These clusters not only can store significantly high amount of hydrogen but also possesses a fast release kinetics of hydrogen. Therefore, these materials should have pretty wide application in the field of hydrogen storage.

\noindent $^*${Please address all correspondence to: Saswata Bhattacharya (saswata@physics.iitd.ac.in)}
\bibliography{hstore}{}
\end{document}